\documentclass[12pt]{article}
\usepackage{epsfig,cite,amsmath}

\def\beq{\begin{equation}}
\def\eeq{\end{equation}}
\def\beqar{\begin{eqnarray}}
\def\eeqar{\end{eqnarray}}
\def\barr#1{\begin{array}{#1}}
\def\earr{\end{array}}
\def\bfi{\begin{figure}}
\def\efi{\end{figure}}
\def\btab{\begin{table}}
\def\etab{\end{table}}
\def\bce{\begin{center}}
\def\ece{\end{center}}

\def\text{\textstyle}

\def\arraystretch{1.4}

\def\al{\alpha}
\def\be{\beta}

\def\de{\delta}

\def\si{\sigma}
\def\Si{\Sigma}

\def\De{\Delta}

\def\refeq#1{\mbox{eq.~(\ref{#1})}}

\def\reffi#1{\mbox{Fig.~\ref{#1}}}

\def\refta#1{\mbox{Table~\ref{#1}}}

\def\citere#1{\mbox{Ref.~\cite{#1}}}

\def\mathswitchr#1{\relax\ifmmode{\mathrm{#1}}\else$\mathrm{#1}$\fi}

\newcommand{\PW}{\mathswitchr W}
\newcommand{\PZ}{\mathswitchr Z}
\newcommand{\PA}{\mathswitchr A}

\newcommand{\PH}{\mathswitchr H}

\newcommand{\Ph}{\mathswitchr h}

\newcommand{\Pb}{\mathswitchr b}

\newcommand{\Pt}{\mathswitchr t}

\def\mathswitch#1{\relax\ifmmode#1\else$#1$\fi}

\newcommand{\MW}{\mathswitch {M_\PW}}

\newcommand{\MZ}{\mathswitch {M_\PZ}}
\newcommand{\MH}{\mathswitch {M_\PH}}
\newcommand{\MHSM}{\mathswitch {M_\PH}^{\rm SM}}
\newcommand{\mHp}{m_{\PH^\pm}}

\newcommand{\mG}{m_{\rm G}}

\newcommand{\mGp}{m_{{\rm G}^\pm}}

\newcommand{\Mb}{\mathswitch {m_\Pb}}
\newcommand{\Mt}{\mathswitch {m_\Pt}}

\newcommand{\mh}{\mathswitch {m_\Ph}}
\newcommand{\mH}{\mathswitch {m_\PH}}
\newcommand{\MA}{\mathswitch {M_\PA}}

\newcommand{\scrs}{{}}
\newcommand{\sw}{\mathswitch {s_{\scrs\PW}}}

\newcommand{\cw}{\mathswitch {c_{\scrs\PW}}}
\newcommand{\sweff}{\sin^2 \theta_{\mathrm{eff}}}

\def\tb{\tan\beta}

\newcommand{\Xt}{X_{\Pt}}

\def\order#1{${\cal O}(#1)$}
\newcommand{\mhmax}{\mh^{\rm max}}

\newcommand{\mt}{\Mt}

\newcommand{\mb}{\Mb}

\newcommand{\Stop}{\tilde{t}}

\newcommand{\tsf}{\theta\kern-.20em_{\tilde{f}}}
\newcommand{\tsfp}{\theta\kern-.20em_{\tilde{f}\prime}}
\newcommand{\tsq}{\theta\kern-.15em_{\tilde{q}}}

\newcommand{\msusy}{M_{\mathrm{SUSY}}}

\newcommand{\gsim}
{\;\raisebox{-.3em}{$\stackrel{\displaystyle >}{\sim}$}\;}

\newcommand{\alps}{\alpha_{\mathrm s}}

\newcommand{\SM}{{\mathrm{SM}}}
\newcommand{\MSSM}{{\mathrm{MSSM}}}

\newcommand{\feh}{{\em FeynHiggs}}

\newcommand{\oaas}{{\cal O}(\alpha\alps)}
\newcommand{\cp}{{\cal CP}}

\newcommand{\twol}{two-loop}
\newcommand{\onel}{one-loop}

\newcommand{\VL}{\left( \begin{array}{c}}
\newcommand{\VR}{\end{array} \right)}
\newcommand{\ML}{\left( \begin{array}{cc}}
\newcommand{\MLd}{\left( \begin{array}{ccc}}
\newcommand{\MLv}{\left( \begin{array}{cccc}}
\newcommand{\MR}{\end{array} \right)}

\newcommand{\tev}{\,\, \mathrm{TeV}}
\newcommand{\gev}{\,\, \mathrm{GeV}}
\newcommand{\mev}{\,\, \mathrm{MeV}}

\newcommand{\BC}{\begin{center}}
\newcommand{\EC}{\end{center}}
\newcommand{\BE}{\begin{equation}}
\newcommand{\EE}{\end{equation}}
\newcommand{\BEA}{\begin{eqnarray}}
\newcommand{\BEAnn}{\begin{eqnarray*}}
\newcommand{\EEA}{\end{eqnarray}}
\newcommand{\EEAnn}{\end{eqnarray*}}
\newcommand{\non}{\nonumber}
\newcommand{\id}{{\rm 1\kern-.12em
\rule{0.3pt}{1.5ex}\raisebox{0.0ex}{\rule{0.1em}{0.3pt}}}}

\def\als{\alpha_s}

\def\draftdate{\relax}
\def\mda{\relax}
\def\mua{\relax}
\def\mla{\relax}
\def\draft{
\def\thtystars{******************************}
\def\sixtystars{\thtystars\thtystars}
\typeout{}
\typeout{\sixtystars**}
\typeout{* Draft mode!
         For final version remove \protect\draft\space in source file
*}
\typeout{\sixtystars**}
\typeout{}
\def\draftdate{\today}
\def\mua{\marginpar[\boldmath\hfil$\uparrow$]%
                   {\boldmath$\uparrow$\hfil}%
                    \typeout{marginpar: $\uparrow$}\ignorespaces}
\def\mda{\marginpar[\boldmath\hfil$\downarrow$]%
                   {\boldmath$\downarrow$\hfil}%
                    \typeout{marginpar: $\downarrow$}\ignorespaces}
\def\mla{\marginpar[\boldmath\hfil$\rightarrow$]%
                   {\boldmath$\leftarrow $\hfil}%
                    \typeout{marginpar:
$\leftrightarrow$}\ignorespaces}
\def\Mua{\marginpar[\boldmath\hfil$\Uparrow$]%
                   {\boldmath$\Uparrow$\hfil}%
                    \typeout{marginpar: $\Uparrow$}\ignorespaces}
\def\Mda{\marginpar[\boldmath\hfil$\Downarrow$]%
                   {\boldmath$\Downarrow$\hfil}%
                    \typeout{marginpar: $\Downarrow$}\ignorespaces}
\def\Mla{\marginpar[\boldmath\hfil$\Rightarrow$]%
                   {\boldmath$\Leftarrow $\hfil}%
                    \typeout{marginpar:
$\Leftrightarrow$}\ignorespaces}
\overfullrule 5pt
\oddsidemargin -15mm
\marginparwidth 29mm
}

\begin{document}
\thispagestyle{empty}

\def\thefootnote{\fnsymbol{footnote}}

\begin{flushright}
DCPT/02/154\\
IPPP/02/77\\
LMU 01/03\\
hep-ph/0301062 \\
\end{flushright}

\mbox{}

\vspace{2cm}

\begin{center}

{\large\sc {\bf Precision Observables in the MSSM:}}

\vspace{0.4cm}

{\large\sc {\bf Status and Perspectives}}%
\footnote{Talk given by G.~Weiglein at SUSY02, June 2002, DESY, Germany}

\vspace{1cm}

{\sc 
S.~Heinemeyer$^{1}$%
\footnote{email: Sven.Heinemeyer@physik.uni-muenchen.de}%
~and G.~Weiglein$^{2}$%
\footnote{email: Georg.Weiglein@durham.ac.uk}
}

\vspace*{1cm}

{\sl
$^1$Institut f\"ur theoretische Elementarteilchenphysik,
LMU M\"unchen, Theresienstr.\ 37, D-80333 M\"unchen, Germany

\vspace*{0.4cm}

$^2$Institute for Particle Physics Phenomenology,\\ 
University of Durham, Durham DH1 3LE, U.K.
}

\end{center}

\vspace*{1.2cm}

\begin{abstract}
The current status of the theoretical predictions for the electroweak
precision observables $\MW$, $\sweff$ and $\mh$ within the MSSM is 
briefly reviewed. The impact of recent electroweak two-loop results to
the quantity $\De\rho$ is analysed and the sensitivity of the
electroweak precision observables to the top-quark Yukawa coupling is
investigated. Furthermore the level of precision necessary
to match the experimental accuracy at the next generation of colliders
is discussed.
\end{abstract}

\def\thefootnote{\arabic{footnote}}
\setcounter{page}{0}
\setcounter{footnote}{0}

\newpage


\section{Electroweak precision observables in the MSSM}

Electroweak precision tests, i.e.\ the comparison of accurate
measurements with predictions of the theory at the quantum level, allow 
to set indirect constraints on unknown parameters of the model under
consideration. 
Within the Standard Model (SM) precision observables like the W-boson
mass, $\MW$, and the effective leptonic weak mixing angle, $\sweff$, allow in
particular to obtain constraints on the Higgs-boson mass of the SM. 
In the Minimal Supersymmetric extension of the SM (MSSM), on the other 
hand, the mass of the lightest $\cp$-even Higgs boson, $\mh$, can be 
predicted in terms of the mass of the $\cp$-odd Higgs boson, $\MA$, and
$\tan\be$, the ratio of the vacuum expectation values of the two Higgs
doublets. Via radiative corrections it furthermore sensitively depends
on the scalar top and bottom sector of the MSSM. Thus, within the MSSM a
precise measurement of $\MW$, $\sweff$ and $\mh$ allows to obtain
indirect information in particular on the parameters of the Higgs and 
scalar top and bottom sector.

The status of the theoretical predictions for $\mh$ within the MSSM has
recently been reviewed in \citere{mhiggsAEC}. The theoretical
predictions, based on the complete one-loop and the 
dominant two-loop results, currently have an uncertainty
from unknown higher-order corrections of about $\pm 3$~GeV, while the
parametric uncertainty from the experimental error of the top-quark mass
presently amounts to about $\pm 5$~GeV.

For the electroweak precision observables within the SM very accurate
results are available. This holds in particular for the prediction for 
$\MW$, where meanwhile all ingredients of the complete two-loop result
are known. The remaining theoretical uncertainties from unknown
higher-order corrections within the SM are estimated to
be~\cite{MWferm2,snowmass,radcor02}
\beq
{\rm SM:} \quad \de \MW^{\rm th} \approx \pm 6 \mev, \quad
\de \sweff^{\rm th} \approx \pm 7 \times 10^{-5}.
\label{eq:unchighord}
\eeq
They are smaller at present than the parametric uncertainties
from the experimental errors of the input parameters $\mt$ and
$\De\al_{\rm had}$. The experimental errors of $\de \mt = \pm
5.1$~GeV~\cite{blueband_s02} and
$\de(\De\al_{\rm had}) = 36 \times 10^{-5}$~\cite{blueband_s02}
induce parametric theoretical uncertainties of
\beqar
\de\mt: \quad
\de \MW^{\rm para} \approx \pm 31 \mev, && 
     \de \sweff^{\rm para} \approx \pm 16 \times 10^{-5}, \non \\
\de(\De\al_{\rm had}): \quad
\de \MW^{\rm para} \approx \pm 6.5 \mev, && 
     \de \sweff^{\rm para} \approx \pm 13 \times 10^{-5}. 
\eeqar
For comparison, the present experimental errors of $\MW$
and $\sweff$ are~\cite{blueband_s02}
\beq
\de \MW^{\rm exp} \approx \pm 34 \mev, \quad
\de \sweff^{\rm exp} \approx \pm 17 \times 10^{-5}.
\label{eq:experr}
\eeq

At one-loop order, complete results for the electroweak precision
observables $\MW$ and $\sweff$ are also known within the
MSSM~\cite{dr1lA,dr1lB}. At the two-loop level, the leading
corrections in $\oaas$ have been obtained~\cite{dr2lA}, which enter via the 
quantity $\De\rho$,
\BE
\De\rho = \frac{\Si_\PZ(0)}{\MZ^2} - \frac{\Si_\PW(0)}{\MW^2} .
\label{delrho}
\EE
It parameterises the leading universal corrections to the electroweak
precision observables induced by
the mass splitting between fields in an isospin doublet~\cite{rho}.
$\Si_{\rm Z,W}(0)$ denote the transverse parts of the unrenormalised Z-
and W-boson self-energies at zero momentum transfer, respectively.
The induced shifts in $\MW$ and $\sweff$ are in leading order given by
(with $1-\sw^2 \equiv \cw^2 = \MW^2/\MZ^2$)
\BE
\de\MW \approx \frac{\MW}{2}\frac{\cw^2}{\cw^2 - \sw^2} \De\rho, \quad
\de\sweff \approx - \frac{\cw^2 \sw^2}{\cw^2 - \sw^2} \De\rho .
\label{precobs}
\EE
For the gluonic corrections, results in $\oaas$ have also been obtained
for the prediction of $\MW$~\cite{dr2lB}. The comparison with the
contributions entering via $\De\rho$ showed that in this case indeed the
full result is well approximated by the $\De\rho$ contribution. Contrary
to the SM case, the \twol\ $\oaas$ corrections turned out to increase
the \onel\ contributions, leading to an enhancement of up to
35\%~\cite{dr2lA}.

Recently the leading \twol\ corrections to 
$\De\rho$ at \order{\al_{\rm t}^2}, \order{\al_{\rm t} \al_{\rm b}}, \order{\al_{\rm b}^2}
have been obtained for the case of a large SUSY scale, 
$\msusy \gg \MZ$~\cite{dr2lal2mh0,dr2lal2}. These contributions involve
the top and bottom Yukawa couplings and
contain in particular corrections proportional to $\mt^4$ and bottom
loop corrections enhanced by $\tb$. 
Since for a large SUSY scale the
contributions from loops of SUSY particles decouple from physical
observables, the leading contributions can be obtained in this case in
the limit where besides the SM particles only the two Higgs doublets of
the MSSM are active. In the following section these results are briefly
summarised.

Comparing the presently available results for the electroweak precision
observables $\MW$ and $\sweff$ in the MSSM with those in the SM, the
uncertainties from unknown higher-order corrections within
the MSSM can be estimated to be at least a factor of two larger than the
ones in the SM as given in \refeq{eq:unchighord}.


\section{Leading electroweak two-loop contributions to $\De\rho$}

The leading contributions of
\order{\al_{\rm t}^2}, \order{\al_{\rm t} \al_{\rm b}} and \order{\al_{\rm b}^2} to $\De\rho$ 
in the limit of a large SUSY scale arise from two-loop diagrams
containing a quark loop and the scalar particles of the two Higgs
doublets of the MSSM, see \citere{dr2lal2}. They 
can be obtained by extracting the
contributions proportional to $y_{\rm t}^2$, $y_{\rm t} y_{\rm b}$ and $y_{\rm b}^2$, where
\BE
y_{\rm t} = \frac{\sqrt{2} \, \mt}{v \, \sin\be}, \quad
y_{\rm b} = \frac{\sqrt{2} \, \mb}{v \, \cos\be} ~.
\label{ytyb}
\EE
The coefficients of these terms can then be evaluated in the gauge-less 
limit, i.e.\ for $\MW, \MZ \to 0$ (keeping $\cw = \MW/\MZ$ fixed).

In this limit the tree-level masses of the charged Higgs boson $H^{\pm}$ 
and the unphysical scalars $G^0$, $G^{\pm}$ are given by
\beq
\mHp^2 = \MA^2, \quad
\mG^2 = \mGp^2 = 0 .
\label{chargedHiggsMW0}
\eeq

Applying the corresponding limit also in the neutral $\cp$-even Higgs
sector would yield for the lightest $\cp$-even Higgs-boson mass
$\mh^2 = 0$ and furthermore $\mH^2 = \MA^2$, $\sin\al = -\cos\be$, 
$\cos\al = \sin\be$, where $\al$ is the mixing angle of the neutral
$\cp$-even states. However, in the SM the limit $\MHSM \to 0$ turned out
to be only a poor approximation of the result for arbitrary $\MHSM$, and
the same feature was found for the limit $\mh \to 0$ within the
MSSM~\cite{dr2lal2mh0,dr2lal2}. Furthermore, the neutral $\cp$-even
Higgs sector is known to receive very large radiative corrections.
Thus, using the tree-level masses in the
gauge-less limit turns out to be a very crude approximation. It
therefore is useful to keep the parameters of the neutral $\cp$-even
Higgs sector arbitrary as far as possible (ensuring a complete
cancellation of the UV-divergences), although the contributions going
beyond the gauge-less limit of the tree-level masses are formally of
higher order. In particular, keeping $\al$ arbitrary is necessary in
order to incorporate non SM-like couplings of the lightest $\cp$-even 
Higgs boson to fermions and gauge bosons.

We first discuss the results for the \order{\al_{\rm t}^2} corrections, which
are by far the dominant subset within
the SM, i.e.\ the \order{\al_{\rm t} \al_{\rm b}} and \order{\al_{\rm b}^2} corrections
can safely be neglected within the SM. The same is true within the MSSM
for not too large values of $\tan\beta$. In this case no further
relations in the neutral $\cp$-even Higgs sector are necessary, i.e.\
the parameters $\mh, \mH$ and $\al$ can be kept arbitrary in the
evaluation of the \order{\al_{\rm t}^2} corrections. For these contributions
also the top Yukawa coupling $y_{\rm t}$ can be
treated as a free parameter, i.e.\ it is not necessary to use
\refeq{ytyb}. This allows to study the sensitivity of the electroweak
precision observables to variations in the top Yukawa coupling.

\begin{figure}[htb!]
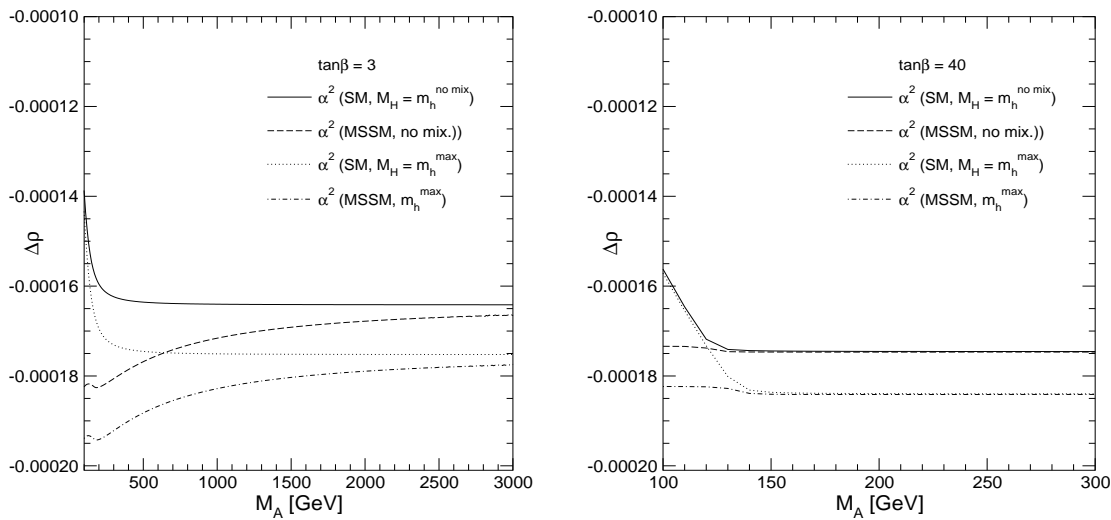

\vspace{1em}
\begin{center}
\mbox{
\epsfig{figure=delrhoMT2Yukfull12b.bw.eps,width=7cm,height=6.8cm}
\hspace{1em}
\epsfig{figure=delrhoMT2Yukfull16.bw.eps,width=7cm,height=6.8cm}
}
\end{center}
\caption[]{
The \order{\al_{\rm t}^2} MSSM contribution to $\De\rho$ in the $\mhmax$ and
the no-mixing scenario is compared with the corresponding SM result with
$\MHSM = \mh$. The value of $\mh$ is obtained in the left (right)
plot from varying $\MA$ from  $100 \gev$ to 3000 (300)~GeV,
while keeping $\tb$ fixed at $\tb = 3$~(40).
}
\label{fig:madep}
\end{figure}

In \reffi{fig:madep} the \order{\al_{\rm t}^2} contribution to $\De\rho$
is shown as a function of $\MA$ for $\tb = 3$ (left plot) and 
$\tb = 40$ (right plot). The values of $\mh$, $\mH$, and $\al$
have been obtained using radiative corrections in the $t/\Stop$ sector
up to two-loop order as implemented in the program
\feh~\cite{feynhiggs}. The MSSM parameters have
been chosen according to the $\mhmax$ and no-mixing scenario benchmark
scenarios~\cite{LHbenchmark}. For comparison, the SM result as given in 
\citere{drSMgf2mt4} is also shown, where for the SM Higgs-boson mass the 
value of $\mh$ has been used.

The \order{\al_{\rm t}^2} MSSM contribution is of \order{10^{-4}}. Its
absolute value is always 
larger than the corresponding SM result. For large values of $\MA$ the
MSSM contribution becomes SM-like. While for $\tb = 3$ this decoupling 
proceeds rather slow and the SM value is reached only for $\MA \gsim 3
\tev$, for $\tb = 40$ the MSSM contribution becomes SM-like already for
quite small $\MA$ values. For very small values of $\MA$, on the other
hand, the behaviour of the SM and the MSSM contributions is very different.
While the SM contribution depends sensitively on $\MHSM$, in the MSSM
for $\MA \gsim 100 \gev$ the dependence on the Higgs boson masses is
much less pronounced, see in particular the right plot of
\reffi{fig:madep}.

The effect of the \order{\al_{\rm t}^2} correction to $\De\rho$ 
on the electroweak precision observables $\MW$ and $\sweff$ amounts to
about $-10 \mev$ for $\MW$ and $+6 \times 10^{-5}$ for $\sweff$, see
\citere{dr2lal2}.
The `effective' change in $\MW$ and $\sweff$ 
in comparison with the corresponding SM result with
the same value of the Higgs-boson mass is significantly smaller. It
amounts up to $-3 \mev$ for $\MW$ and $+2 \times 10^{-5}$ for $\sweff$. 
The effective change goes to zero for large $\MA$ as expected from the
decoupling behaviour.

\begin{figure}[htb!]
\begin{center}
\epsfig{figure=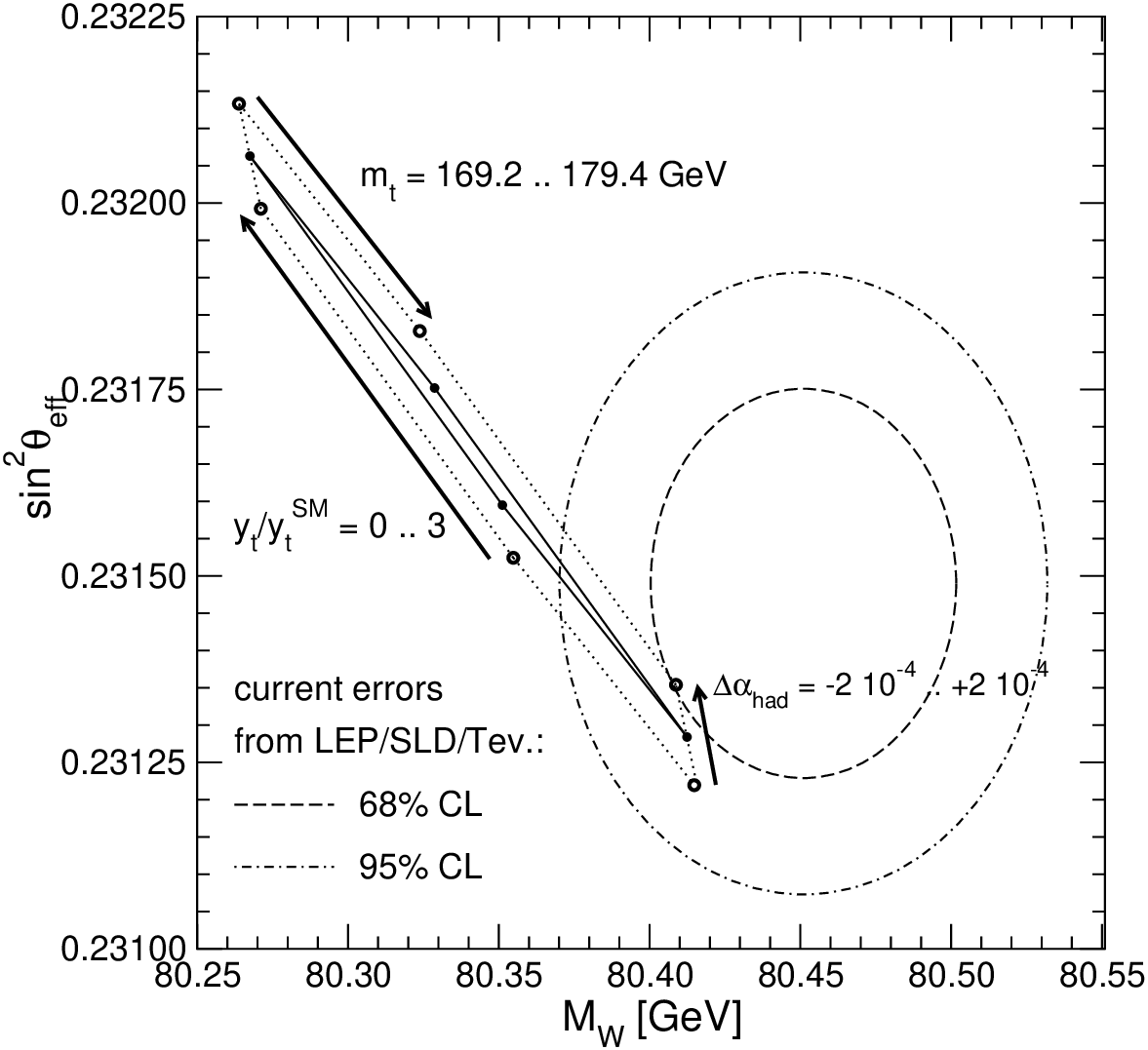,width=11.5cm,height=8.5cm} \\[1em]
\epsfig{figure=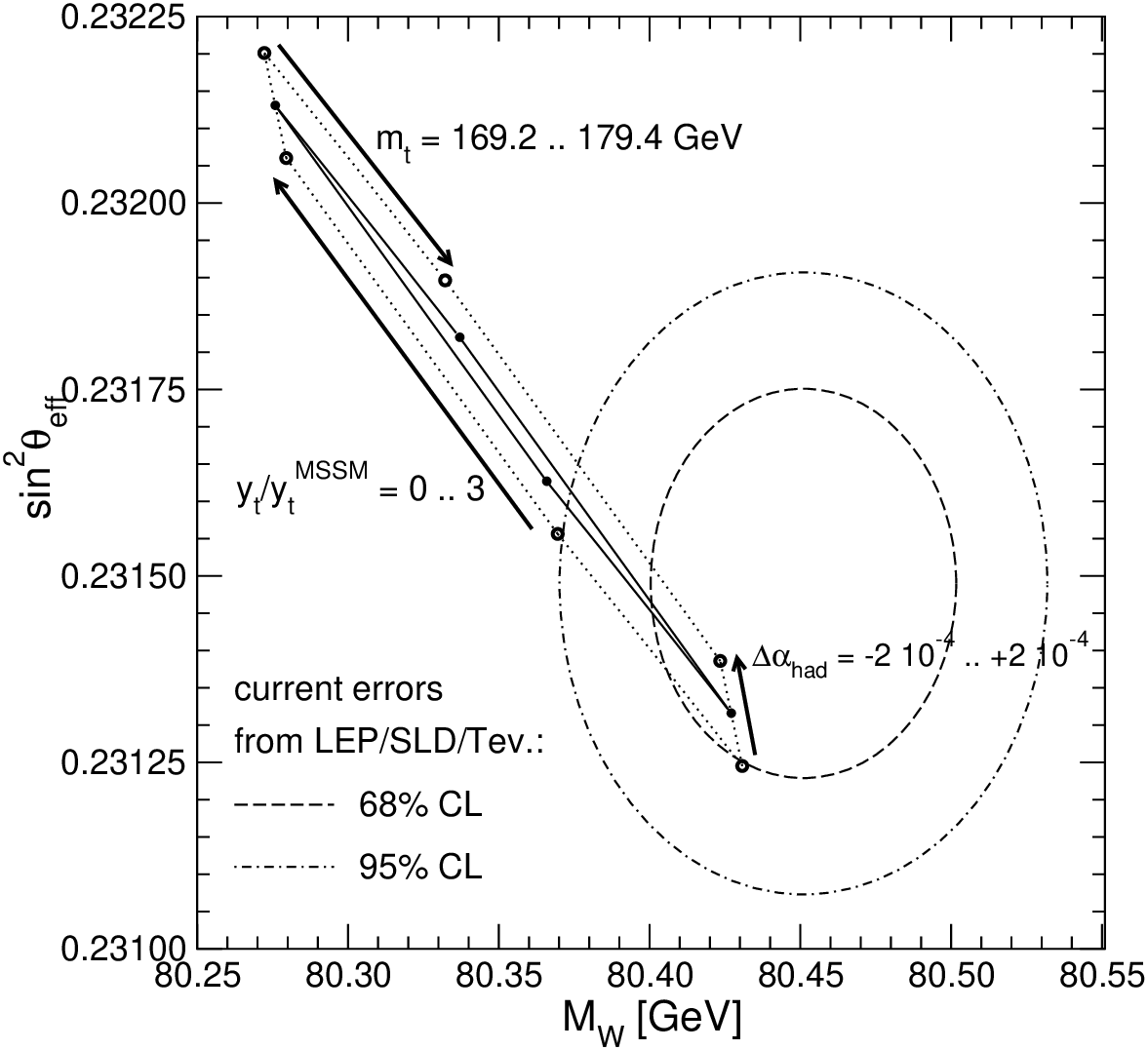,width=11.5cm,height=8.5cm} 
\end{center}
\caption[]{
The effect of scaling the top Yukawa 
coupling in the SM (upper plot) and the MSSM (lower plot) for the
observables $\MW$ and $\sweff$ is shown in comparison with
the current experimental precision.  
The variation with $\mt$ and $\De\al_{\rm had}$ is shown within their
current experimental errors. 
For the SM evaluation, $\MHSM$ has
been set to the conservative value of $\MHSM = 114 \gev$ (see text). 
For the MSSM evaluation the parameters are $\msusy = 1000 \gev$, 
$\Xt = 2000 \gev$, $\MA = 175 \gev$, $\tb = 3$ and $\mu = 200 \gev$,
resulting in $\mh \approx 114 \gev$.
}
\label{fig:canyoutellme}
\end{figure}

\smallskip
Since the top Yukawa coupling enters the predictions for the
electroweak precision observables at lowest order in the perturbative
expansion at \order{\al_{\rm t}^2}, these contributions allow to study the 
sensitivity of the precision observables to this coupling. 
This sensitivity is indicated in \reffi{fig:canyoutellme}, where the 
top Yukawa coupling in the SM and the MSSM is treated as if it were a
free parameter. For simplicity, the top Yukawa coupling entering the SM
contribution is scaled compared to its SM value in the following way,
\BE
y_{\rm t} = x \, y_{\rm t}^{\SM} , \quad 0 \leq x \leq 3 ,
\EE
and analogously in the MSSM. The corresponding variation of the
theoretical prediction for $\MW$ and $\sweff$ 
is compared with the current experimental precision. 
The allowed 68\% and 95\% C.L.\ contours are indicated in the figure.

For the evaluation of $\MW$ and $\sweff$ in the SM and the MSSM,
the complete \onel\ results as well as the leading
\twol\ \order{\al\als} and \order{\al_{\rm t}^2} corrections
have been taken into account.
Since the SM prediction deviates more from the experimental central
value for increasing values of $\MHSM$, $\MHSM = 114
\gev$~\cite{LEPHiggsSM} has been chosen in the figure as a conservative
value.
The current $1\si$ uncertainties in $\mt$ and $\De\al_{\rm had}$ are
also taken into account, as indicated in the plots. 
Varying the SM top Yukawa coupling (upper plot) yields an upper
bound of $y_{\rm t} < 1.3 \, y_{\rm t}^{\SM}$ for $\mt = 174.3 \gev$ 
and of $y_{\rm t} < 2.2 \, y_{\rm t}^{\SM}$ for $\mt = 179.4 \gev$, both at the
95\% C.L. These relatively strong bounds are of course related to the
fact that the theory prediction in the SM shows some deviation from
the current experimental central value. 

The lower plot of \reffi{fig:canyoutellme} shows the analogous
analysis in the MSSM for one particular example of SUSY parameters.
We have chosen a large value of $\msusy$, $\msusy = 1000 \gev$,
in order to justify the approximation of neglecting the
\order{\al_{\rm t}^2} contributions from SUSY loops. The other parameters are
$\Xt = 2000 \gev$, $\MA = 175 \gev$, $\tb = 3$ and $\mu = 200 \gev$,
resulting in $\mh \approx 114 \gev$ (for comparison with the SM
case). The SUSY contributions to $\MW$ and $\sweff$ lead to a somewhat
better agreement between the theory prediction and experiment and
consequently to somewhat weaker bounds on $y_{\rm t}$. In this example we
find $y_{\rm t} < 1.7 \, y_{\rm t}^{\MSSM}$ for $\mt = 174.3 \gev$ and 
$y_{\rm t} < 2.5 \, y_{\rm t}^{\MSSM}$ for $\mt = 179.4 \gev$, both at the 95\%~C.L. 

\smallskip
Including besides the \order{\al_{\rm t}^2} corrections also the \order{\al_{\rm t}
\al_{\rm b}} and \order{\al_{\rm b}^2} corrections to $\De\rho$ into the analysis
requires further symmetry relations as a consequence of the
SU(2)~structure of the fermion doublet.
Within the Higgs boson sector it is necessary, besides
using \refeq{chargedHiggsMW0}, also to use the relations for the
heavy $\cp$-even Higgs-boson mass and the Higgs
mixing angle,
\BE
\mH^2 = \MA^2 , \quad
\sin\al = -\cos\be , \quad
\cos\al = \sin\be .
\label{heavyHiggsMW0}
\EE
On the other hand, $\mh$ can be kept as a
free parameter. The couplings of the lightest $\cp$-even Higgs boson
to gauge bosons and SM fermions,
however, become SM-like, once the mixing angle relations,
\refeq{heavyHiggsMW0}, are used.
Corrections enhanced by
$\tb$ thus arise only from the heavy Higgs bosons, while the
contribution from the lightest $\cp$-even Higgs boson resembles the SM
one. 
Furthermore, the Yukawa couplings can no longer be treated as free
parameters, i.e.\ \refeq{ytyb} has to be employed (and the corresponding 
relations for the SM contribution), which ensures that
the Higgs mechanism governs the Yukawa couplings.

\begin{figure}[htb!]
\begin{center}
\mbox{
\epsfig{figure=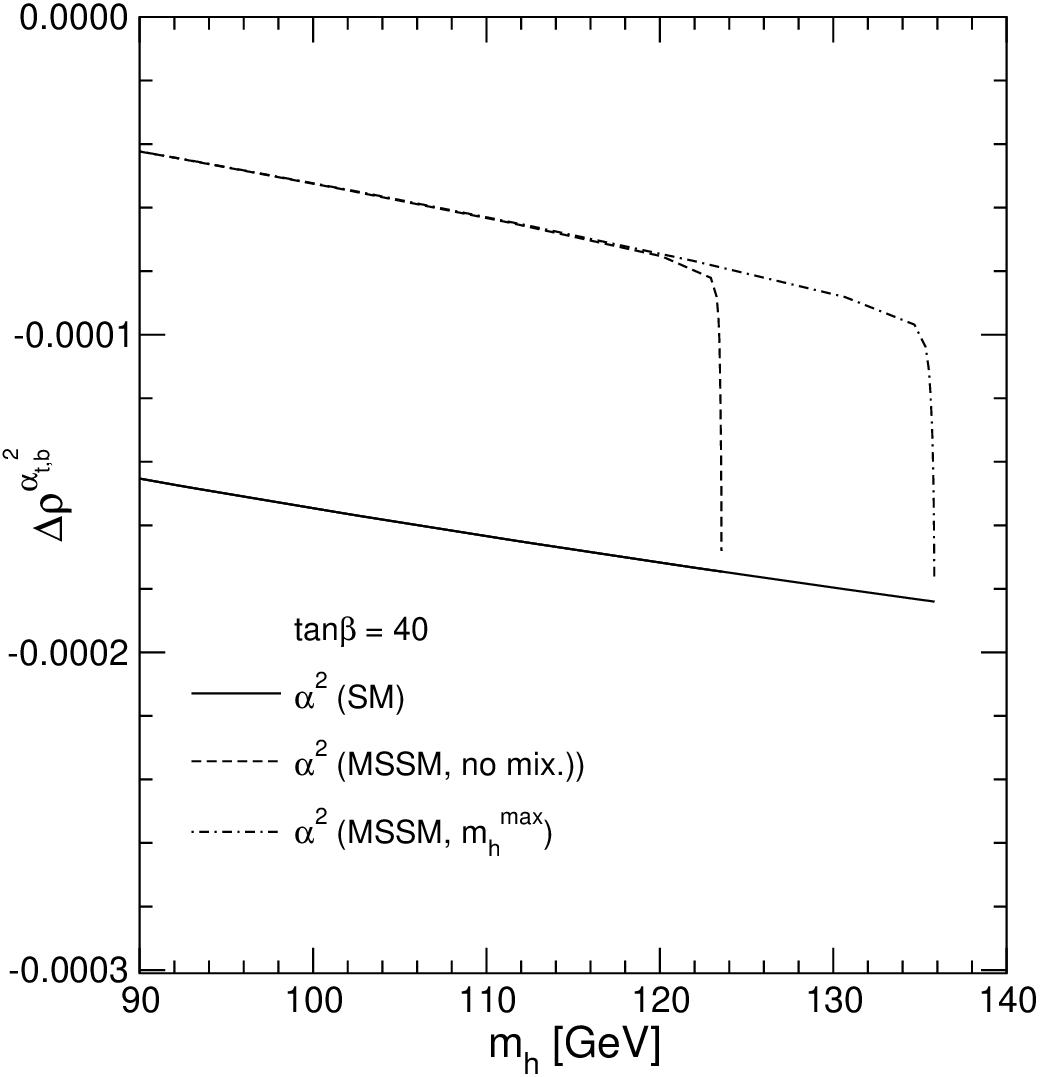,width=7cm,height=6.8cm} 
\hspace{1em}
\epsfig{figure=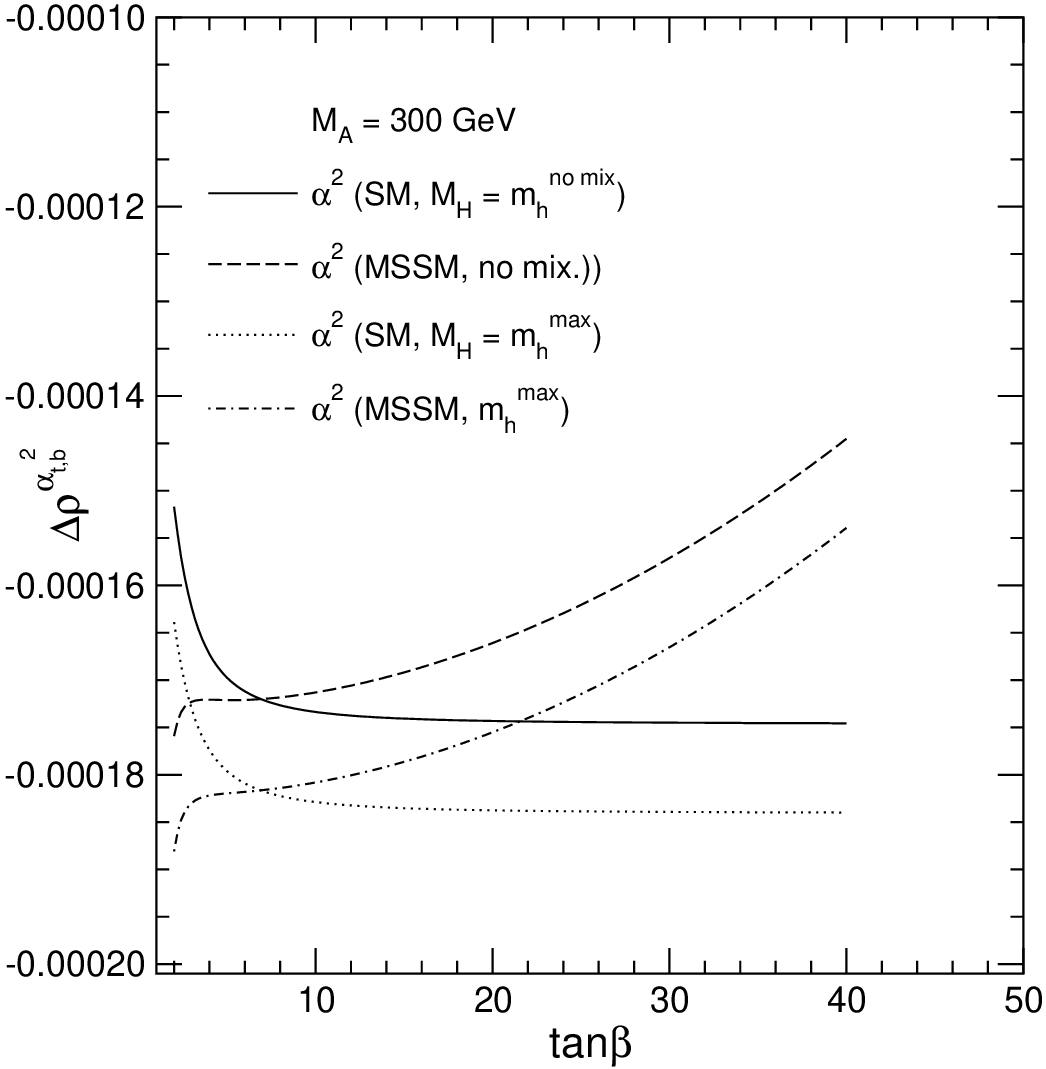,width=7cm,height=6.8cm} 
}
\end{center}
\caption[]{
The \order{\al_{\rm t}^2}, \order{\al_{\rm t} \al_{\rm b}}, and \order{\al_{\rm b}^2} MSSM
contribution to $\De\rho$ in the $\mhmax$ and the no-mixing scenario
is compared with the corresponding SM result with $\MHSM = \mh$. 
In the left plot $\tb$ is fixed to $\tb = 40$, while $\MA$
is varied from $50 \gev$ to $1000 \gev$. In the right plot $\MA$ is
set to $300 \gev$, while $\tb$ is varied. The bottom-quark mass is set
to $\mb = 4.25 \gev$.
}
\label{fig:delrhotb}
\end{figure}

In \reffi{fig:delrhotb} the result for the \order{\al_{\rm t}^2},
\order{\al_{\rm t} \al_{\rm b}}, and \order{\al_{\rm b}^2} MSSM contributions to 
$\De\rho$ is shown in the $\mhmax$ and the no-mixing scenario, 
compared with the corresponding SM result with $\MHSM = \mh$. 
In the left plot $\tb$ is fixed to $\tb = 40$ and $\MA$ is varied from
$50 \gev$ to $1000 \gev$. In the right plot $\MA$
is fixed to $\MA = 300 \gev$, while $\tb$ is varied. 
For large $\tb$ the \order{\al_{\rm t} \al_{\rm b}} and \order{\al_{\rm
b}^2}
contributions yield a significant effect from the heavy Higgs bosons
in the loops, entering with the other sign than the \order{\al_{\rm t}^2}
corrections, while the contribution of the lightest Higgs boson is
SM-like. As one can see in \reffi{fig:delrhotb}, for large $\tb$ the
MSSM contribution to $\De\rho$ is smaller than the SM value. For large
values of $\MA$ the SM result is recovered. 
The effective change in the predictions for the precision observables
from the \order{\al_{\rm t} \al_{\rm b}} and \order{\al_{\rm b}^2} 
corrections can
exceed the one from the \order{\al_{\rm t}^2} corrections. It can amount up
to $\de\MW \approx +5 \mev$ and $\de\sweff \approx -3\times 10^{-5}$
for $\tb = 40$.


\section{Prospects for the next generation of colliders}

The experimental determination of the
electroweak precision observables will improve in the future at the
LHC and even more at a LC with a GigaZ option, see 
\citere{snowmass} for a detailed discussion.
The prospects for the measurements of $\sweff$, $\MW$, $\mt$
and the Higgs-boson mass are summarised in
\refta{tab:futureunc}~\cite{snowmass,gigaz}. 

\begin{table}[htb]
\renewcommand{\arraystretch}{1.5}
\begin{tabular}{|c||c||c|c|c|c||c|c|}
\cline{2-8} \multicolumn{1}{c||}{}
& now & Tev.\ Run~IIA & Run~IIB & Run~IIB$^*$ & LHC & ~LC~  & GigaZ \\
\hline\hline
$\de\sweff(\times 10^5)$ & 17   & 78   & 29   & 20   & 14--20 & (6)  & 1.3  \\
\hline
$\de\MW$ [MeV]           & 34   & 27   & 16   & 12   & 15   & 10   & 7      \\
\hline
$\de\mt$ [GeV]           &  5.1 &  2.7 &  1.4 &  1.3 &  1.0 &  0.2 & 0.13   \\
\hline
$\de\MH$ [MeV]            &  --- &  --- &
                 \multicolumn{2}{c|}{${\cal O}(2000)$} &  100 &   50 &   50 \\
\hline
\end{tabular}
\renewcommand{\arraystretch}{1}
\caption{Current and anticipated future experimental uncertainties for 
$\sweff$, $\MW$, $\mt$ and $\MH$, see \citere{snowmass} for a detailed 
discussion and further references.}
\label{tab:futureunc}
\end{table}

The improvement in the measurement of $\mt$ and prospective future
improvements in the determination of $\De\al_{\rm had}$ will furthermore
lead to a drastic reduction of the parametric theoretical 
uncertainties induced by
the experimental errors of the input parameters. In \refta{tab:para}
the current parametric theoretical uncertainties are compared with the
prospective situation after several years of LC running, for which we
have assumed $\de\mt^{\rm future} = 0.13 \gev$. For the uncertainty in
$\De\al_{\rm had}$ at this time we have assumed $\de(\De\al_{\rm had})^{\rm
future} = 5 \times 10^{-5}$. At this level of accuracy also the
experimental uncertainty of the Z-boson mass, $\de\MZ = 2.1 \mev$, will
be non-negligible, which is not
expected to improve in the foreseeable future. In the scenario discussed
here the parametric theoretical uncertainty in $\MW$ will be
significantly smaller than the prospective experimental error, while for
$\sweff$ the parametric theoretical uncertainty will be slightly larger
than the prospective experimental error.

\begin{table}[htb]
\renewcommand{\arraystretch}{1.5}
\BC
\begin{tabular}{|c||c|c||c|c||}
\cline{2-5} 
\multicolumn{1}{c||}{} & 
\multicolumn{2}{c||}{now} &
\multicolumn{2}{c|}{future} \\ 
\cline{2-5} 
\multicolumn{1}{c||}{} & 
$\de\MW^{\rm para}$ [MeV] & $\de\sweff^{\rm para} (\times 10^5)$ &
$\de\MW^{\rm para}$ [MeV] & $\de\sweff^{\rm para} (\times 10^5)$ \\ 
\hline\hline
$\de\mt$ & 31 & 16 & 1 & 0.5 \\ 
\hline
$\de\al_{\rm had}$ & 6.5 & 13 & 1 & 1.8 \\ 
\hline
$\de\MZ$ & 2.5 & 1.4 & 2.5 & 1.4 \\ 
\hline
\end{tabular}
\EC
\renewcommand{\arraystretch}{1}
\caption{Current and anticipated future parametric uncertainties for
$\sweff$ and $\MW$. For the experimental error of the top-quark mass, 
$\de\mt$, we have assumed 
$\de\mt^{\rm today} = 5.1 \gev$, $\de\mt^{\rm future} = 0.13 \gev$,
for $\de(\De\al_{\rm had})$ we have used
$\de(\De\al_{\rm had})^{\rm today} = 3.6 \times 10^{-4}$, 
$\de(\De\al_{\rm had})^{\rm future} = 5 \times 10^{-5}$, while the
experimental error of the Z-boson mass, $\de\MZ = 2.1 \mev$, is not
expected to improve in the foreseeable future.
}
\label{tab:para}
\end{table}

Similarly as for $\MW$ and $\sweff$, the parametric theoretical
uncertainty of the lightest $\cp$-even Higgs-boson mass in the MSSM 
induced by the experimental error of the top-quark mass will drastically
improve. The prospective accuracy for $\mt$ at the LC will reduce this 
uncertainty to the level of \order{100}~MeV.

In order not to be limited by the theoretical uncertainties from unknown
higher-order corrections, electroweak precision tests after several
years of LC running will require to reduce the latter uncertainties to the
level of about $\pm 3$~MeV for $\MW$ and $\pm 1 \times 10^{-5}$ for
$\sweff$. Achieving this level of accuracy within the MSSM or further
extensions of the SM will clearly require a lot of effort. For the
prediction of the lightest $\cp$-even Higgs-boson mass in the MSSM an
improvement of the theoretical uncertainties from unknown higher-order
corrections by about a factor of 30 compared to the present situation 
will be necessary in order to match the experimental accuracy achievable
at the next generation of colliders.


\subsection*{Acknowledgements}
G.W.\ thanks the organisers of SUSY02 for the invitation and the very
pleasant atmosphere at the conference. 
This work has been supported by the European Community's Human
Potential Programme under contract HPRN-CT-2000-00149 Physics at
Colliders.




\begin{thebibliography}{00}  

\bibitem{mhiggsAEC} 
G.~Degrassi, S.~Heinemeyer, W.~Hollik, P.~Slavich and G.~Weiglein, 
hep-ph/0212020.

\bibitem{MWferm2}
A.~Freitas, W.~Hollik, W.~Walter and G.~Weiglein,
{\em Nucl.\ Phys.} {\bf B 632} (2002) 189.

\bibitem{snowmass}
U.~Baur, R.~Clare, J.~Erler, S.~Heinemeyer, D.~Wackeroth, G.~Weiglein
and D.~R.~Wood,
hep-ph/0111314,
in {\it Proc. of the APS/DPF/DPB Summer Study on the Future of Particle
Physics (Snowmass 2001)} eds.\ R.~Davidson and C.~Quigg.

\bibitem{radcor02}
A.~Freitas, S.~Heinemeyer and G.~Weiglein,
hep-ph/0212068.

\bibitem{blueband_s02}
M.W.~Gr\"unewald,
hep-ex/0210003,
talk given at ICHEP02, Amsterdam, July 2002.

\bibitem{dr1lA} R.~Barbieri and L. Maiani,  
                {\em Nucl. Phys.} {\bf B 224} (1983) 32; \\
                C.~S.~Lim, T.~Inami and N.~Sakai, 
                {\em Phys. Rev.} {\bf D 29} (1984) 1488; \\
                E.~Eliasson, 
                {\em Phys. Lett.} {\bf B 147} (1984) 65; \\
                Z.~Hioki, 
                {\em Prog. Theo. Phys.} {\bf 73} (1985) 1283; \\
                J.~A.~Grifols and J.~Sola, 
                {\em Nucl. Phys.} {\bf B 253} (1985) 47; \\
                B.~Lynn, M.~Peskin and R.~Stuart, 
                CERN Report 86-02, p. 90; \\
                R.~Barbieri, M.~Frigeni, F.~Giuliani and H.E.~Haber, 
                {\em Nucl. Phys.} {\bf B 341} (1990) 309; \\   
                M.~Drees and K.~Hagiwara, 
                {\em Phys. Rev.} {\bf D 42} (1990) 1709.

\bibitem{dr1lB} M.~Drees, K.~Hagiwara and A.~Yamada, 
                {\em Phys. Rev.} {\bf D 45} (1992) 1725; \\ 
                P.~Chankowski, A.~Dabelstein, W.~Hollik, W.~M\"osle, 
                S.~Pokorski and J.~Rosiek, 
                {\em Nucl. Phys.} {\bf B 417} (1994) 101;\\ 
                D.~Garcia and J.~Sol\`a, 
                {\em Mod. Phys. Lett.} {\bf A 9} (1994) 211.


\bibitem{dr2lA} A.~Djouadi, P.~Gambino, S.~Heinemeyer, W.~Hollik,
                C.~J\"unger and G.~Weiglein, 
                {\em Phys. Rev. Lett.} {\bf 78} (1997) 3626,
                hep-ph/9612363;
                {\em Phys. Rev.} {\bf D 57} (1998) 4179,
                hep-ph/9710438.

\bibitem{rho} M.~Veltman, 
              {\em Nucl. Phys.} {\bf B 123} (1977) 89. 

\bibitem{dr2lB} 
                S.~Heinemeyer, PhD thesis, 
                Universit\"at Karlsruhe, Shaker Verlag, Aachen 1998,
                see {\tt www-itp.physik.uni-karlsruhe.de/prep/phd/};\\
                G.~Weiglein, 
                hep-ph/9901317;\\
                S.~Heinemeyer, W.~Hollik and G.~Weiglein,
                {\em in preparation}.
                
\bibitem{dr2lal2mh0} S.~Heinemeyer and G.~Weiglein, hep-ph/0102317.

\bibitem{dr2lal2} S.~Heinemeyer and G.~Weiglein,
                  {\em JHEP} {\bf 10} (2002) 072,
                  hep-ph/0209305.

\bibitem{feynhiggs} S.~Heinemeyer, W.~Hollik and G.~Weiglein, {\em
                    Comp. Phys. Comm.} {\bf 124} 2000 76,
                    hep-ph/9812320;
                    hep-ph/0002213;\\
                    M.~Frank, S.~Heinemeyer, W.~Hollik and G.~Weiglein,
                    hep-ph/0202166.\\
                    The code is accessible via
                    {\tt www.feynhiggs.de} .

\bibitem{LHbenchmark} M.~Carena, S.~Heinemeyer, C.~Wagner and G.~Weiglein,
                      to appear in {\em Eur. Phys. Jour.} {\bf C}, 
                      hep-ph/0202167.

\bibitem{drSMgf2mt4} R.~Barbieri, M.~Beccaria, P.~Ciafaloni, G.~Curci 
                     and A.~Vicere,
                     {\em Nucl. Phys.} {\bf B 409} (1993) 105;\\
                     J.~Fleischer, F.~Jegerlehner and O.V.~Tarasov,
                     {\em Phys. Lett.} {\bf B 319} (1993) 249.

\bibitem{LEPHiggsSM} [LEP Higgs working group], LHWG Note/2002-01,\\
                     {\tt http://lephiggs.web.cern.ch/LEPHIGGS/papers/}.

\bibitem{gigaz} J.~Erler, S.~Heinemeyer, W.~Hollik, G.~Weiglein 
                and P.M.~Zerwas,
                {\em Phys. Lett.} {\bf B 486} (2000) 125,
                hep-ph/0005024.

\end{thebibliography}
\end{document}